\documentclass{ws-procs975x65}

\begin{document}

{\hfill UNF-Astro-3-1-10C}

{\hfill \ } 

\title{SENSITIVITY OF THE FERMI DETECTORS TO GAMMA-RAY BURSTS FROM EVAPORATING PRIMORDIAL BLACK HOLES (PBHS)}

\author{T. N. UKWATTA$^{1,2}$, J. H. MACGIBBON$^{3*}$, W. C. PARKE$^1$, K. S. DHUGA$^1$, S. RHODES$^3$, A. ESKANDARIAN$^1$, N. GEHRELS$^2$, L. MAXIMON$^1$, and D. C. MORRIS$^{1,2}$}

\address{$^1$The George Washington University,\\
Washington, D.C. 20052, USA}

\address{$^2$ NASA Goddard Space Flight Center,\\
Greenbelt, MD 20771, USA}

\address{$^3$Department of Physics, University of North Florida,\\
Jacksonville, FL 32224, USA\\
$^*$E-mail: jmacgibb@unf.edu}

\begin{abstract}
Primordial Black Holes (PBHs), which may have been created in the
early Universe, are predicted to be detectable by their Hawking
radiation. The Fermi Gamma-ray Space Telescope observatory
offers increased sensitivity to the gamma-ray bursts produced by
PBHs with an initial mass of $\sim  5\times 10^{14}$ g expiring
today. PBHs are candidate progenitors of unidentified Gamma-Ray
Bursts (GRBs) that lack X-ray afterglow. We propose spectral lag,
which is the temporal delay between the high and low energy
pulses, as an efficient method to identify PBH evaporation events
with the Fermi Large Area Telescope (LAT).
\end{abstract}


\bodymatter

\section{Introduction}\label{aba:sec1}
The formation of Primordial Black Holes (PBHs) has been postulated
in many theories of the early Universe (for a recent review see
Ref.~\refcite{C05}). Black holes of mass $M_{\rm bh}$ continually
emit Hawking radiation~\cite{H} with a temperature of $T_{\rm bh}=
1.06\ {\rm GeV}/\left( M_{\rm bh}/10^{13}\ {\rm g} \right)$ in the
form of all available fundamental particle species. The emitted
particles decay quickly on astrophysical timescales into $\gamma$,
$\nu_{e,\mu,\tau}$, $\bar{\nu}_{e,\mu,\tau}$, $p$, $\bar{p}$,
$e^+$ and $e^-$. PBHs with an initial mass\cite{MCP} of $M_*\sim
5\times 10^{14}$ g should be expiring today with a burst of high
energy particles including gamma-rays. The current upper limit on
the number expiring today per volume per unit time is\cite{MC}
\begin{equation}
R\lesssim 10^{-7}\eta_{\rm \, local}\, \rm{pc^{-3} \, yr^{-1}}
\label{aba:eq1}
\end{equation}
where $\eta_{\rm \, local}$ is the density enhancement of PBHs in
the local region. Typically $\eta_{\rm \, local}$ is $\sim 10^6$
(for clustering in the Galactic halo) or larger. Such PBH bursts
may be detectable by the Fermi Gamma-ray Space Telescope
observatory's Large Area Telescope (LAT). Conversely,
non-detection by the LAT may lead to tighter bounds on the PBH
distribution.

\section{PBH burst as seen by Ideal Detector}
In the standard model\cite{MCP}\,, the total number of photons
emitted per second by a $T_{\rm bh}\sim 0.3 - 100$ GeV black hole
scales as\cite{MW}
\begin{equation}
\dot{N}_{\rm bh \, \gamma} \simeq 1.4\times
10^{29}\left(\frac{T_{\rm bh}}{\rm TeV}\right)^{1.6}\rm{\ s}^{-1}.
\label{aba:eq2}
\end{equation}
The number of photons per second per unit area reaching the Earth
from a PBH bursting at a distance $d$ from the Earth is
\begin{equation}
F_{\rm bh}=\frac{\dot{N}_{\rm bh \, \gamma}}{4\pi d^2}.
\label{aba:eq3}
\end{equation}
Let us assume an ideal detector of effective area $A_{\rm \, eff}$
which can detect every photon that falls on it. (If the detector
is non-ideal then the efficiency can be incorprated into the value
of $A_{\rm \, eff}$.) If the detector requires $X$ photons over
time $t$ to distinguish an incoming event as a burst, then to
detect a burst we require $F_{\rm bh}A_{\rm \, eff}t\geq X$. That
is the PBH must be closer than
\begin{equation}
d_{\rm \, D}\simeq\frac{2.6\times 10^{-2}}{\sqrt{X}}
\left(\frac{T_{\rm bh}}{\rm TeV}\right)^{0.8} \left(\frac{A_{\rm
\, eff}}{{\rm m}^2}\right)^{1/2}\left(\frac{t}{\rm
min}\right)^{1/2}\ {\rm pc}. \label{aba:eq4}
\end{equation}
What $T_{\rm bh}$ maximizes the chance of detection?  The remaining
lifetime\cite{M2} of a PBH of temperature $T_{\rm bh}$ is $\tau_{\rm evap}\simeq 7.4\times 10^3
/\left(T_{\rm bh}/{\rm TeV}\right)^{3} f$ s where $f\left(T_{\rm bh}\right)$ weights the
number of emitted species. (The remaining lifetime of a 300 GeV, 1 TeV, or 5 TeV black hole is 1 hour, 1 minute, or 1 second, respectively.) Taking $t = \tau_{\rm evap}$, a PBH will be detected by the ideal detector if it is closer than
\begin{equation}
d_{\rm \, D}\simeq\frac{0.03}{\sqrt{X}} \left(\frac{T_{\rm
bh}}{\rm TeV}\right)^{-0.7} \left(\frac{A_{\rm \, eff}}{{\rm
m}^2}\right)^{1/2}\ {\rm pc}. \label{aba:eq5}
\end{equation}
Thus the detectability is maximized for the lowest $T_{\rm bh}$
black hole visible above the background and/or by using the
longest detector exposure time. For a detector of angular
resolution $\Omega$ to resolve the PBH above the gamma-ray
background $F_\gamma$, we also require that $F_{\rm bh}A_{\rm \,
eff}\geq F_\gamma \Omega A_{\rm \, eff}$. The PBH will be resolved
above the observed (EGRET) extragalactic background\cite{S}
\begin{equation}
\frac{dF_\gamma}{dE}\simeq 1.4\times 10^{-6} \left(\frac{E}{\rm
GeV}\right)^{-2.1}\ {\rm cm}^{-2}\ {\rm GeV}^{-1}\ {\rm s}^{-1}\
{\rm sr}^{-1} \label{aba:eq6}
\end{equation}
at energy $E$ by the ideal detector if the PBH is closer than
\begin{equation}
d_{\rm \, R}\simeq 0.03\left(\frac{\Omega}{\rm sr
}\right)^{-1/2}\left(\frac{E}{\rm
GeV}\right)^{0.55}\left(\frac{T_{\rm bh}}{\rm TeV}\right)^{0.8}\
{\rm pc} \label{aba:eq7}
\end{equation}
and $E$ is less than the average energy\cite{MW} of the PBH
photons $\overline{E}_{\gamma}\approx 10 \left(T_{\rm bh}/{\rm
TeV}\right)^{0.5}$ GeV. The isotropic diffuse gamma-ray
background, which is an upper limit on the extragalactic
background, recently measured\cite{A} by the LAT at mid-Galactic
latitudes is consistent with the earlier EGRET measurements Eq.
(\ref{aba:eq6}), although the extragalactic component
may\cite{SMR} be weaker by up to a factor of 2.

For a given detector, the scanned volume of space is then $V_{\rm
bh} = \left(\omega_{\rm A}/{\rm sr}\right)d_{\rm \, S}^{\,3}/3$
where $\omega_{\rm A}$ is the detector acceptance angle (field of
view) and $d_{\rm \, S} = \min (d_{\rm \, D}, d_{\rm \, R})$.
Extensive air shower arrays characteristically have $A_{\rm \,
eff}\gtrsim 10^4\ {\rm m}^2$,  large $\omega_{\rm A}$ and small
$\Omega$ but very high threshold energy (typically $\gtrsim 10$
TeV) and hence are background-limited, while atmospheric Cerenkov
detectors\cite{AB} characteristically have $A_{\rm \, eff}\gtrsim
200\ {\rm m}^2$ and small $\Omega$ but high threshold energy
(typically $\gtrsim 100$ GeV although the Whipple SGARFACE
system\cite{LKS} has a threshold of 100 MeV) and very small
$\omega_{\rm A}$ ($\lesssim 10^{-2}$ sr). In contrast, the Fermi
LAT has\cite{AT} a smaller $A_{\rm \, eff}\sim 0.8\ {\rm m}^2$ but
large $\omega_{\rm A} \sim 2.4$ sr, finer source position angular
resolution ($0.3 - 2\, '$), low energy thresholds (down to 20 MeV),
good time resolution and is essentially background-free with
respect to burst sensitivity. Additionally, most of the photons
emitted by expiring $T_{\rm bh}\lesssim 1$ TeV PBHs are in the LAT
energy range (20 MeV - 300 GeV).

\section{Conclusions}
The Fermi LAT offers greater sensitivity to local PBH bursts
than ground-based detectors. We have proposed\cite{T} spectral lag
measurements (the temporal delay between high and low energy
pulses) of the incoming light curve in two different energy bands
as a method to identify PBH bursts. A PBH burst arriving at the
detector will exhibit positive to negative evolution with
increasing energy because the black hole temperature and
$\overline{E}_{\gamma}$ increase over time as the black hole loses
mass. Because spectral lag measurements require counts in only two
energy bands, and not the full spectrum, spectral lag can be
measured even for weak events that last for very short time
scales. Work is in progress to calculate quantitative values for
the PBH spectral lags for the characteristics of the Fermi LAT.


\begin{thebibliography}{9}
\bibitem{C05} B.~J. Carr, in {\it Inflating Horizons of Particle Astrophysics and Cosmology}, (Universal Academy Press Inc, Tokyo, 2006).
\bibitem{H} S.~W. Hawking, {\em Nature} {\bf 248}, 30 (1974);  S.~W. Hawking, {\em Commun.  Math. Phys.} {\bf 43}, 199 (1975); S.~W. Hawking, {\em Phys. Rev. D} {\bf 13}, 191 (1976).
\bibitem{MCP} J.~H. MacGibbon, B.~J. Carr, and D.~N. Page, {\em Phys. Rev. D} {\bf 78}, 064043 (2008).
\bibitem{MC} J.~H. MacGibbon and B.~J. Carr, {\em Astrophys. J.} {\bf 371}, 447 (1991).
\bibitem{MW} J.~H. MacGibbon and B.~R. Webber, {\em Phys. Rev. D} {\bf 41}, 3052 (1990).
\bibitem{M2} J.~H. MacGibbon, {\em Phys. Rev. D} {\bf 44}, 376 (1991).
\bibitem{S} P. Shreekumar {\it et al.}, {\em Astrophys. J.} {\bf 494}, 523 (1998).
\bibitem{A} A.~A. Abdo {\it et al.}, {\em Astrophys. J.} {\bf 703}, 1249 (2009).
\bibitem{SMR} A.~W. Strong, I.~V. Moskalenko, and O. Reimer, {\em Astrophys. J.} {\bf 613}, 956 (2004).
\bibitem{AB} L.~A. Antonelli {\it et al.}, {\it The Next Generation of Cherenkov Telescopes. A White Paper for the Italian National Institute for Astrophysics (INAF)} arXiv:0906.4114.
\bibitem{LKS} S. LeBohec, F. Krennrich, and G. Sleege {\em Astropart. Phys.} {\bf 23}, 235 (2005).
\bibitem{AT} W.~B. Atwood {\it et al.}, {\it The Large Area Telescope on the Fermi Gamma-ray Space Telescope Mission} arXiv:0902.1089.
\bibitem{T} T.~N. Ukwatta {\it et al.}, {\em AIP Conf. Proc.} {\bf 1133}, 440 (2009).
\end{thebibliography}
\end{document}